\documentclass[conference]{IEEEtran}
\usepackage{amsfonts}

\usepackage{enumerate}

\usepackage{amsmath,amsthm}
\usepackage[noend]{algpseudocode}
\usepackage{float}
\usepackage{hyperref}
\usepackage{color}
\usepackage{makeidx}
\usepackage{bbm}
\usepackage{graphicx}
\usepackage{lipsum}
\usepackage{soul}
\usepackage{tabularx}
\usepackage{dsfont}
\usepackage{multirow}

\usepackage{amsfonts}
\usepackage{times}
\usepackage{graphicx}
\usepackage{latexsym}
\usepackage{dsfont}
\usepackage{amssymb}
\usepackage{amsmath}
\usepackage{cite}
\usepackage{verbatim}
\usepackage{subfigure}

\newcommand{\figref}[1]{{Fig.}~\ref{#1}}
\newcommand{\tabref}[1]{{Table}~\ref{#1}}

\def\bb0{{\mathbb{0}}}

\def\bb{{\mathbf{b}}}

\def\b0{{\mathbf{0}}}

\def\sf0{{\mathsf{0}}}

\usepackage{epstopdf}

\newcommand{\fref}[1]{{Fig.}~\ref{#1}}

\newcommand{\subto}{\operatorname{s.t.}}

\begin{document}
\title{Predicting Future CSI Feedback For Highly-Mobile Massive MIMO Systems}
\author{Yu Zhang, Ahmed Alkhateeb, Pranav Madadi, Jeongho Jeon, Joonyoung Cho, Charlie Zhang \thanks{Yu Zhang and Ahmed Alkhateeb are with Arizona State University (Email: y.zhang, alkhateeb@asu.edu), Pranav Madadi, Jeongho Jeon, Joonyoung Cho and Charlie Zhang are with Samsung Research America.}}
\maketitle

\begin{abstract}

Massive multiple-input multiple-output (MIMO) system is promising in providing unprecedentedly high data rate. To achieve its full potential, the transceiver needs complete channel state information (CSI) to perform transmit/receive precoding/combining. This requirement, however, is challenging in the practical systems due to the unavoidable processing and feedback delays, which oftentimes degrades the performance to a great extent, especially in the high mobility scenarios.
In this paper, we develop a deep learning based channel prediction framework that proactively predicts the downlink channel state information based on the past observed channel sequence.
In its core, the model adopts a 3-D convolutional neural network (CNN) based architecture to efficiently learn the temporal, spatial and frequency correlations of downlink channel samples, based on which accurate channel prediction can be performed.
Simulation results highlight the potential of the developed learning model in extracting information and predicting future downlink channels directly from the observed past channel sequence, which significantly improves the performance compared to the sample-and-hold approach, and mitigates the impact of the dynamic communication environment.

\end{abstract}

\section{Introduction} \label{intro}
Massive multiple-input multiple-output (MIMO) technique plays a central role in 5G and beyond for its capability of providing unprecedentedly high data rate by leveraging the degree of freedom in the spatial domain \cite{Larsson2014}. However, achieving such potential requires the acquisition of accurate channel state information (CSI), which is challenging in most of the practical systems. This is especially the case in a frequency division duplex (FDD) system where the channel estimated in the uplink (UL) cannot be used for downlink (DL) transmission, since channel reciprocity does not hold due to the different carrier frequencies. In addition to the huge training overhead associated with the massive MIMO system, another noticeable phenomenon that significantly impacts the system performance is called \emph{channel aging} \cite{Truong2013}. To put it in simple words, channel aging models the channel variation from the time point when it is learned to the actual time when it is used. This captures the inherent time-varying channel characteristic and the processing delay in a practical system. To mitigate such degradation caused by channel aging, future channel prediction or channel tracking is required, where the overall objective is to proactively predict what could be the actual channel state at the time it is being used, based on available observed channels. This turns out to be possible as wireless channel is essentially a function of its surrounding environment and hence the channel samples are correlated over time \cite{Alrabeiah2019}. Therefore, in this paper, we investigate the problem of predicting future channels by exploiting their temporal correlations with the past observed channel samples for a downlink massive MIMO system.

\textbf{Contribution:}
In this paper, we develop a novel 3-D convolutional neural network (CNN) based deep learning framework for predicting the future CSI in a massive MIMO-OFDM system.
Given the huge dimension of the channel matrices in such system, we leverage the 3-D kernels of the convolutional layer to explore the correlations across frequency and spatial dimensions, while the temporal correlations between channel samples are captured by using the additional ``\emph{channel}'' dimension of a CNN.
Besides, the proposed learning model is enhanced with a residual architecture for extracting deeper yet crucial features that are hardly learned by shallow networks while keeping the overall training experience efficient.
The generalization capability of the model is improved by training the network on a mixed-speed channel dataset, where the channel sequences are collected under different mobilities.
The extensive simulation evaluations on a 3GPP channel model highlight the capability of the proposed solution in accurately predicting future channels, which brings significant improvement over sample-and-hold approach.

\textbf{Prior work:}
The effects of channel aging in massive MIMO systems were first studied in \cite{Truong2013}. The authors of \cite{Zhao2018} proposed a time-varying channel tracking method for massive MIMO systems based on spatial-temporal basis expansion model. \cite{Han2019} utilized the delay and angular reciprocity between the uplink and downlink frequency bands to reduce the amount of unknown parameters to be estimated in the downlink. There are also some machine learning based channel estimation solutions. For example, \cite{Arnold2019} introduced a neural network based scheme for extrapolating downlink CSI from observed uplink CSI. \cite{Yuan2019} first used neural networks to identify the channel aging pattern, then adopted an autoregressor to forecast CSI. Our proposed solution is novel in that we utilize a pure 3-D CNN based deep learning framework with residual architecture to directly predict the future channel in the frequency and spatial domain, based on the observed past channels.



\begin{figure}[t]
	\centering
	\includegraphics[width=\linewidth]{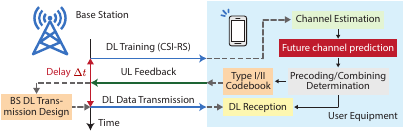}
	\caption{The considered system model where a multi-antenna base station is communicating with a multi-antenna user equipment. To perform DL data transmission, the DL CSI should be first estimated and sent back to the base station by the user equipment. However, due to the processing delay, the channel varies by the time base station transmits the data.}
	\label{sys_model}
\end{figure}

\section{System Model} \label{sec:System}

We consider a frequency division duplex (FDD) downlink single-cell system where a base station (BS) with $N_t$ antennas is communicating with a single user equipment (UE) with $N_r$ antennas. We assume an OFDM based system with a total number of $K$ sub-bands. Since channel reciprocity does not hold in a FDD system, the BS needs to rely on the channel state information (CSI) feedback from UE to determine its downlink data transmission scheme.

\textbf{Downlink Training and Uplink Feedback:}
In the downlink training stage, the BS sends downlink pilots to the UE, based on which the downlink channel is estimated. In this paper, we assume that the UE is able to estimate the downlink channel perfectly, although channel estimation is always challenging in the practical applications.
We assume that the channel matrix at the $k$-th sub-band is $\mathbf{H}_k\in\mathbb{C}^{N_r\times N_t}$. The UE will then feed back the downlink CSI to the BS. However, as can be seen, directly sending back $\mathbf{H}_k$ for all the sub-bands will incur huge feedback overhead.
Therefore, to reduce the feedback overhead, a common practice is that the UE first calculates the desired beamforming vectors based on the estimated downlink channel, and then feeds back those beamforming vectors to the BS. To be more specific, the downlink beamforming vector at the $k$-th sub-band can be obtained by using the right singular vector\footnote{Here, without loss of generality, we assume a single layer transmission, which in practice, is informed by rank indicator also fed back by UE.} of $\mathbf{H}_k$, i.e., $\mathbf{f}_k=\left[\mathbf{V}_k\right]_{:, 1}$, where $\mathbf{H}_k=\mathbf{U}_k\boldsymbol{\Lambda}_k\mathbf{V}_k^H$ is the singular value decomposition (SVD) of $\mathbf{H}_k$. Finally, the set of beamforming vectors $\mathcal{F}=\left\{\mathbf{f}_1, \mathbf{f}_2, \dots, \mathbf{f}_K\right\}$ will be sent to the BS.

It is worth mentioning that, in practice, such feedback overhead oftentimes is still too expensive to afford by the system. Therefore, in the current 3GPP release, those beamforming vectors will further be approximated by using Type I/II codebooks \cite{TS_38_214} (which is essentially the discrete Fourier transform (DFT) vector basis), and those coefficients will finally be sent out, as shown in \figref{sys_model}. As this is not the focus of this paper, we assume that the beamforming vectors designed by the UE can be perfectly re-constructed at the BS and used for subsequent downlink data transmission.

\textbf{Downlink Data Transmission:}
Based on the uplink feedback, the BS then transmits downlink data using the beamforming vectors provided by the UE. Accordingly, the UE will use the optimal combining vectors for reception, i.e., the left singular vector of the estimated $\mathbf{H}_k$, that is, $\mathbf{w}_k=\left[\mathbf{U}_k\right]_{:, 1}$. Therefore, the sum rate across all the sub-bands is given by
\begin{equation}\label{sum-rate}
  R = \sum_{k=1}^{K}\log_2\left(1 + \frac{P_x}{\sigma_n^2}\left|\mathbf{w}_k^H \mathbf{H}_k \mathbf{f}_k\right|^2\right),
\end{equation}
where $P_x$ is the average transmit power, $\sigma_n^2$ is the noise power, and $\mathbf{H}_k$ is assumed to incorporate the large-scale fading factor.

\section{Problem Formulation} \label{sec:Problem}

As can be seen from the above described system model, the \emph{ideally} achieved system performance \eqref{sum-rate} relies on multiple factors. One of the most important assumption is that the beamforming vector at the BS and the combining vector at the UE indeed correspond to the channel when the BS transmits downlink data. In practice, however, there is always a delay between the time point when the channel is estimated and the time when actually downlink transmission happens. In other words, by the time the BS starts transmitting data with the beamforming vector fed back by the UE, the channel already changes. Moreover, such delay is always in the level of milliseconds. This could be quite problematic in high mobility scenarios where the channel changes very fast, which degrades the system performance to a great extent. Therefore, proactive channel prediction is necessary to mitigate such impact.

In this paper, we investigate the channel prediction problem in a downlink FDD massive MIMO system. The objective is to maximize the throughput of \textbf{future} downlink transmission, based on the knowledge of past and current channels. Formally, the problem can be cast as
\begin{align}\label{Prob}
 \max_{\left\{\widehat{\mathbf{f}}_k, \widehat{\mathbf{w}}_k\right\}_{k=1}^K} & \hspace{6pt} \sum_{k=1}^{K}\log_2\left(1 + \frac{P_x}{\sigma_n^2}\left|\widehat{\mathbf{w}}_k^H \mathbf{H}_k^{(t+1)} \widehat{\mathbf{f}}_k\right|^2\right), \\
 \subto ~ & \|\widehat{\mathbf{f}}_k\|_2 \le 1, ~ \|\widehat{\mathbf{w}}_k\|_2 \le 1, \forall k \\
 ~ & \hspace{2pt} \mathbf{H}_k^{(t^\prime)} ~ \text{until time $t^\prime=t$ are available},
\end{align}
%
%
where $\widehat{\mathbf{f}}_k$ and $\widehat{\mathbf{w}}_k$ are the $k$-th sub-band beamforming and combining vectors, $\mathbf{H}_k^{(t+1)}$ is the downlink channel matrix at the $k$-th sub-band and time $t+1$. Therefore, the problem is essentially to design $\widehat{\mathbf{f}}_k$ and $\widehat{\mathbf{w}}_k$ for the transmission at time $t+1$ based on the observed channels up to time $t$.
%

It is worth pointing out that there are multiple ways to proactively design such transmission/reception schemes.
In this paper, we follow the method of first predicting the channel at time $t+1$, based on which transmission and reception schemes can be derived based on SVD. Therefore, the core of the problem becomes proactively predicting the channel at the transmission time, i.e., time $t+1$, based on the previous and current estimated channels, i.e., up to time $t$. To put it in formal form, we want to \textbf{learn a channel predictor} that takes in the past channel observations and outputs the future channel, that is
\begin{equation}\label{ch-pred-func}
  \widehat{\mathbb{H}}^{(t+1)} = f\left(\mathbb{H}^{(t)}, \mathbb{H}^{(t-1)}, \dots, \mathbb{H}^{(t-(L-1))}; \boldsymbol{\Theta}\right),
\end{equation}
where $f$ is the desired channel predictor parameterized by $\boldsymbol{\Theta}$ and $L$ is the number of past channel observations used for prediction. We use $\mathbb{H}^{(t)} \in \mathbb{C}^{K \times N_r \times N_t}$, for simplicity, to denote the 3-D downlink channel matrix in the frequency and spatial domain at time $t$. Therefore, we have
\begin{equation}\label{3dchannel}
  \left[\mathbb{H}^{(t)}\right]_{k,:,:} = \mathbf{H}_k^{(t)}.
\end{equation}
Based on the predicted channel $\widehat{\mathbb{H}}^{(t+1)}$, the beamforming and combining vectors $\widehat{\mathbf{f}}_k, \widehat{\mathbf{w}}_k, \forall k$ can be correspondingly constructed by using SVD.
Hence, the problem becomes finding $\boldsymbol{\Theta}$ that yields accurate channel prediction, by using the past $L$ channel observations.
As the problem essentially features learning a function, we resort to the powerful neural networks to be the channel predictor.


\begin{figure}[t]
	\centering
	\includegraphics[width=\linewidth]{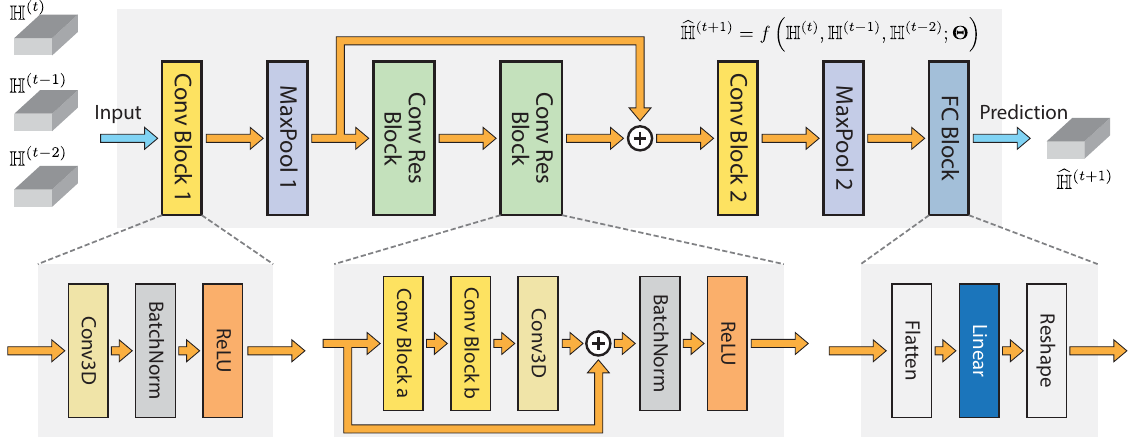}
	\caption{The proposed 3-D convolutional neural network based deep learning model for future channel prediction. The schematic shows an example of $L=3$.}
	\label{dl_model}
\end{figure}

\section{Proposed Machine Learning Solution} \label{sec:Sol}

In this section, we describe the detailed architecture of the proposed deep learning model which is based on convolutional neural networks, and discuss the loss function used for training the network.

\subsection{Network Architecture}

The building block of the proposed deep learning model is a 3-D convolutional layer, where the 3-D kernel is utilized to extract the features across the frequency and spatial dimensions (transmit and receive antennas). Different features are then added up together to form \emph{one of} the input feature maps of the next layer, where the temporal correlation can be captured. As can be seen in \figref{dl_model}, there are two different blocks building on top of the 3-D convolutional layer. The first one is called convolutional block, which basically includes extra normalization layer and activation layer. Another one named convolutional residual block is further built on top of the convolutional block, where a residual architecture is adopted. The introducing of residual architecture is mainly used for extracting deeper features that are hardly found by shallow networks while keeping the training experience efficient. Finally, a fully-connected (FC) block is employed to reshape the output to have the desired dimension. The detailed network architecture can be found at \tabref{NetParam}, where $B$ stands for the batch size, and the input dimension of the FC block, $X$, can be calculated as follow
\begin{equation}\label{Xdim}
  X = 2\times\left\lfloor\frac{N_r}{1}\right\rfloor\times\left\lfloor\frac{N_t}{2}\right\rfloor\times\left\lfloor\frac{K}{4}\right\rfloor,
\end{equation}
where $\lfloor\cdot\rfloor$ is the floor function. It is worth mentioning that the main purpose of introducing the second max-pooling layer is to reduce the number of parameters in the last FC layer, as normally, $N_t$ and $K$ are relatively large.

\begin{table*}[t]
\caption{Parameters of the 3-D CNN residual network}
\centering
\begin{tabular}{c|c|cc}
  \hline
  \hline
  \multicolumn{2}{c|}{\textbf{Module}} & \textbf{Parameter} & \textbf{Value} \\
  \hline
  \multicolumn{2}{c|}{\multirow{2}{*}{\textbf{Conv Block 1}}} & Dimension of input \& output & $(B, 2L, N_r, N_t, K)$ \& $(B, 4L, N_r, N_t, K)$ \\
  \multicolumn{2}{c|}{{}} & Kernel size \& Padding \& Stride & $(3, 7, 5)$ \& $(1, 3, 2)$ \& (1, 1, 1) \\
  \hline
  \multicolumn{2}{c|}{\textbf{MaxPool 1}} & Kernel size \& Padding \& Stride & $(3, 3, 3)$ \& $(1, 1, 1)$ \& (1, 1, 1) \\
  \hline
  \multirow{6}{*}{\textbf{Conv Res Block}} & \multirow{2}{*}{\textbf{Conv Block a}} & Dimension of input \& output & $(B, 4L, N_r, N_t, K)$ \& $(B, 8L, N_r, N_t, K)$ \\
                                                                        &  & Kernel size \& Padding \& Stride & $(3, 7, 5)$ \& $(1, 3, 2)$ \& (1, 1, 1) \\
                                 \cline{2-4}
                                 & \multirow{2}{*}{\textbf{Conv Block b}} & Dimension of input \& output & $(B, 8L, N_r, N_t, K)$ \& $(B, 16L, N_r, N_t, K)$ \\
                                                                 &  & Kernel size \& Padding \& Stride & $(3, 7, 5)$ \& $(1, 3, 2)$ \& (1, 1, 1) \\
                                 \cline{2-4}
                                 & \multirow{2}{*}{\textbf{Conv3D}} & Dimension of input \& output & $(B, 16L, N_r, N_t, K)$ \& $(B, 4L, N_r, N_t, K)$ \\
                                                           &  & Kernel size \& Padding \& Stride & $(3, 7, 5)$ \& $(1, 3, 2)$ \& (1, 1, 1) \\
  \hline
  \multicolumn{2}{c|}{\multirow{2}{*}{\textbf{Conv Block 2}}} & Dimension of input \& output & $(B, 4L, N_r, N_t, K)$ \& $(B, 2, N_r, N_t, K)$ \\
  \multicolumn{2}{c|}{{}} & Kernel size \& Padding \& Stride & $(3, 7, 7)$ \& $(1, 3, 3)$ \& (1, 1, 1) \\
  \hline
  \multicolumn{2}{c|}{\textbf{MaxPool 2}} & Kernel size \& Padding \& Stride & $(1, 2, 4)$ \& $(0, 0, 0)$ \& $(1, 2, 4)$ \\
  \hline
  \multicolumn{2}{c|}{\textbf{FC Block}} & Dimension of input \& output & $(B, X)$ \& $(B, 2\times N_r\times N_t\times K)$ \\
  \hline
  \hline
\end{tabular}
\label{NetParam}
\end{table*}

\subsection{Training Loss Function}

As the target of the neural network based channel predictor $f_{\boldsymbol{\Theta}}$ is to predict the future channels, we pose our learning problem as \textbf{a regression problem} conducted in a supervised learning fashion. Furthermore, we employ mean squared error (MSE) as the training loss function. For each predicted channel sample, the loss function is defined as
\begin{align}\label{loss}
  \mathcal{L}&\left( \mathbb{H}^{(t+1)}, \widehat{\mathbb{H}}^{(t+1)} \right) = \left\| \mathbb{H}^{(t+1)} - \widehat{\mathbb{H}}^{(t+1)} \right\|_F^2, \\
  = & \left\| \mathbb{H}^{(t+1)} - f\left(\{\mathbb{H}^{(q)}\}_{q=t-(L-1)}^t; \boldsymbol{\Theta}\right) \right\|_F^2,
\end{align}
where $\mathbb{H}^{(t+1)}$ is the ground true future channel, $\widehat{\mathbb{H}}^{(t+1)}$ is the prediction of the network, and $\|\cdot\|_F$ stands for the Frobenius norm of a matrix, which can be straightforwardly generalized to a 3-D matrix. For the considered 3-D channel matrix of shape $K\times N_r\times N_t$, we have
\begin{equation}\label{3d-fnorm}
  \|\mathbb{H}\|_F^2 = \sum_{k=1}^{K}\|\mathbf{H}_k\|_F^2.
\end{equation}

The objective of learning is to find a set of model parameters that minimizes the loss averaged over the whole dataset, which can be expressed as
\begin{equation}\label{objLr}
  \min_{\boldsymbol{\Theta}}\frac{1}{N}\sum_{n=1}^{N}\mathcal{L}\left( \mathbb{H}_n, \widehat{\mathbb{H}}_n \right),
\end{equation}
where $N$ is the size of the dataset, the details of which will be given in the next section, and $\boldsymbol{\Theta}$ is the model parameters to be optimized.

\section{Experimental Setup} \label{sec:Exp}

In this section, we describe the adopted channel model, dataset and the training parameters used in our simulations.

\subsection{Channel Model and Dataset}

In our simulations, we consider a massive MIMO-OFDM system where the base station has $N_t=32$ antennas, the user equipment has $N_r=4$ antennas, and there are $K=52$ resource blocks.
The channel coefficients are generated by a system level simulator from Samsung Research America, which follows 3GPP Urban Micro (UMi) channel model with a central frequency of 2.1 GHz and channel bandwidth 20 MHz. Both line-of-sight (LOS) and non-line-of-sight (NLOS) scenarios are considered. We assume a periodic CSI-RS configuration with a periodicity of 5 ms. The mobility of the UE depends on the specific dataset.

%

\textbf{Uni-Speed Dataset:}
Based on the above parameters, for each user, we generate a long channel sequence with a total length of $Q$ channel samples under a certain speed.
It is worth noting that to ensure an efficient learning process, the generated channel sequence is first normalized. This is found to be very useful in the previous work \cite{Zhang2020Deep}. In this work, we adopt the average power normalization method. To be more specific, we first calculate the average power of each \emph{channel element} as follow
\begin{equation}\label{avgPower}
  P = \frac{1}{Q\cdot KN_rN_t}\sum_{n=1}^{Q}\left\|\mathbb{H}_n\right\|_F^2.
\end{equation}
And then, the obtained average channel element power will be used to normalize all the channel samples
\begin{equation}\label{normCh}
  \mathbb{H}_n \doteq \frac{1}{\sqrt{P}}\mathbb{H}_n, ~ \forall n = 1, 2, \dots, Q.
\end{equation}
Such normalized channel sequence is then truncated into segments with length $L+1$. Therefore, we have a total number of $Q-L$ short channel sequence segments, which forms our dataset, that is
\begin{equation}\label{dataset}
  \mathcal{D} = \left\{\left(\mathbb{H}^{t-(L-1)}, \dots, \mathbb{H}^{(t-1)}, \mathbb{H}^{(t)}; \mathbb{H}^{(t+1)}\right)_{t=L}^{Q-1}\right\},
\end{equation}
where, for each data sample, the first $L$ channel matrices are input of the network and the last channel matrix is used as desired output, i.e., training target.

\textbf{Mixed-Speed Dataset:}
Furthermore, instead of assuming that the UE has a constant speed, which is rarely the case in practice, we also consider a more realistic scenario where the UE changes its speed over time. To this end, we generate another kind of dataset that is under a range of UE speed, e.g., 30 km/h to 50 km/h. To be more specific, we first generate multiple channel sequences under different UE speeds. Then, the channel sequence generated under each UE speed will be truncated into short segments as before. It is worth noting that those channel sequences will first go through a similar normalization pre-processing step as performed in the uni-speed dataset before being truncated. Finally, all those channel segments (under different UE speeds) are \emph{mixed} together to form the ultimate mixed-speed dataset. As can be seen, the mixed-speed dataset contains more dynamics than a uni-speed dataset, which helps improve the generalization capability of the channel prediction model.

\subsection{Model Training and Testing}

With the network architecture and datasets described above, the 3-D CNN based channel predictor is then trained and tested over training and testing datasets respectively. The training is performed using PyTorch with an
NVIDIA Quadro RTX 6000 GPU. The detailed hyper-parameters of training are summarized in \tabref{TrParam}.

\begin{table}[t]
\caption{Hyper-parameters for model training}
\centering
\begin{tabular}{c|c}
  \hline
  \hline
  \textbf{Parameter} & \textbf{value} \\
  \hline
  Batch size & 512 \\
  Number of epochs & 300 \\
  Optimizer & Adam \\
  Initial learning rate & $1\times10^{-3}$ \\
  Learning rate schedule & $0.1$@$\left[100, 200, 250\right]$ \\
  Training / Testing & 70\% / 30\% \\
  \hline
  \hline
\end{tabular}
\label{TrParam}
\end{table}

\begin{figure*}[t]
	\centering
	\includegraphics[width=.9\linewidth]{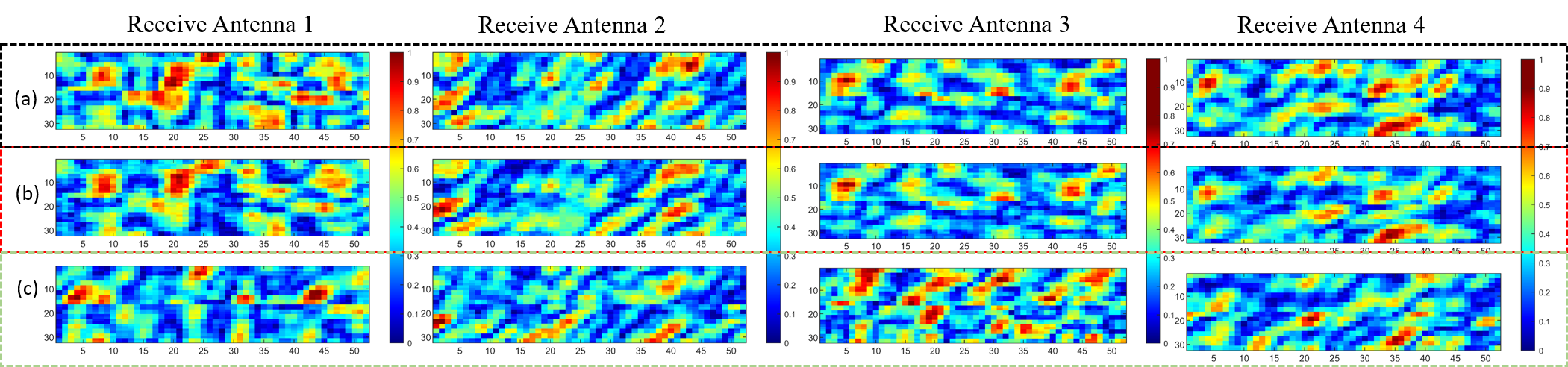}
	\caption{Heat map of the channels at each receive antenna for: (a) The true channel at time $t+1$, i.e., $\mathbb{H}^{(t+1)}$, (b) the predicted channel by the proposed deep learning model, i.e., $\widehat{\mathbb{H}}^{(t+1)}$, and (c) the channel obtained by using sample and hold approach, i.e., channel at the previous time step $t$, $\mathbb{H}^{(t)}$.}
	\label{simu_fig_1}
\end{figure*}

\subsection{Performance Evaluation Metric} \label{svd}

We adopt two different metrics to evaluate the accuracy of the predicted channels. One is for the predicted raw CSI, i.e., the 3-D channel matrix $\widehat{\mathbb{H}}$ in the frequency and spatial domain. Another is for the beamforming vector constructed based on the predicted raw channel.

\textbf{Evaluating the predicted raw CSI:}
We use normalized MSE (NMSE) as the metric to measure the prediction accuracy of the raw CSI, which is defined as
\begin{equation}\label{nmse}
  \mathsf{NMSE}(\mathbb{H}, \widehat{\mathbb{H}}) = \frac{\|\mathbb{H} - \widehat{\mathbb{H}}\|_F^2}{\|\mathbb{H}\|_F^2}.
\end{equation}
It is worth mentioning that oftentimes, the decibel value of NMSE is used, which can be converted from \eqref{nmse} through $\mathsf{NMSE}_{\mathrm{dB}}(\mathbb{H}, \widehat{\mathbb{H}})=10\log_{10}\mathsf{NMSE}(\mathbb{H}, \widehat{\mathbb{H}})$.

\textbf{Evaluating the predicted beamforming vector:}
As the sub-band beamforming vectors are what will finally be fed back to the base station, they are more closely related to the final system performance, e.g., achievable rate or throughput. Therefore, we consider using cosine similarity as the metric to evaluate their accuracy, which turns out to be more suitable than the NMSE. The cosine similarity of two complex vectors $\mathbf{f}$ and $\widehat{\mathbf{f}}$ can be defined as
\begin{equation}\label{cossim}
  \rho(\mathbf{f}, \widehat{\mathbf{f}}) = \frac{|\mathbf{f}^H\widehat{\mathbf{f}}|}{\|\mathbf{f}\|_2\|\widehat{\mathbf{f}}\|_2}.
\end{equation}
As can be seen from the above definition, the cosine similarity between two vectors are independent of their magnitudes. It essentially captures how close they are in term of direction.
Also, as both ground true beamforming vector $\mathbf{f}$ and the beamforming vector obtained based on the predicted channel, i.e., $\widehat{\mathbf{f}}$, are with unit norm, essentially we have $\rho(\mathbf{f}, \widehat{\mathbf{f}})=|\mathbf{f}^H\widehat{\mathbf{f}}|$.

\subsection{Baseline Solutions} \label{baselines}

We compare the performance of the proposed solution with the sample-and-hold (SH) approach, and leave the comparison to other existing approaches for future work.

\textbf{Sample-and-Hold}: As the name suggests, the SH approach stores the most recent estimated channel samples and always sends them to the base station, which will be used for DL transmission. Specifically, in our simulations, we assume that the base station will design the DL transmission scheme based on $\mathbb{H}^{(t)}$, while the actual downlink channel has changed to $\mathbb{H}^{(t+1)}$.
As can be seen, the SH approach essentially ignores the inevitable transmission delay between the time point when the channel is estimated and the time point when the actual DL transmission happens, or assumes that period that the channel stays static is longer than the processing delay. Therefore, the performance of SH approach is overly dependent on how fast the channel is changing as well as the transmission delay, i.e., the difference between $\mathbb{H}^{(t)}$ and $\mathbb{H}^{(t+1)}$.


\section{Simulation Results} \label{sec:Simu}

In this section, we evaluate the performance of our proposed deep learning based future channel prediction approach for both uni-speed and mixed-speed datasets as described before.

\subsection{Results for the Uni-Speed Dataset}

\begin{figure}[t]
	\centering
	\includegraphics[width=3in]{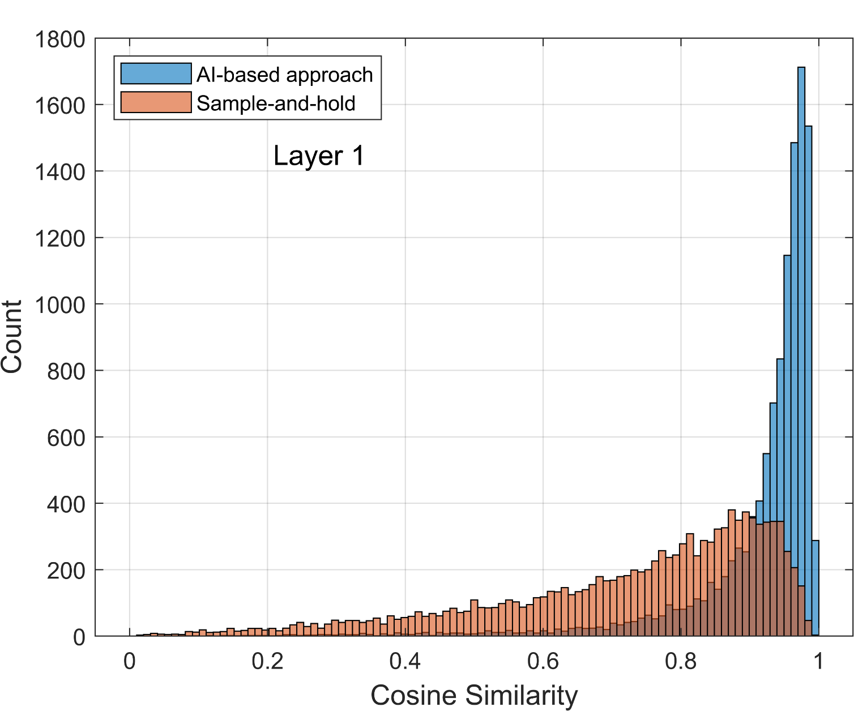}
	\caption{The histogram of the cosine-similarity performance of the AI-based approach and the sample-and-hold approach.}
	\label{simu_fig_2}
\end{figure}

\begin{figure}[t]
	\centering
	\includegraphics[width=3in]{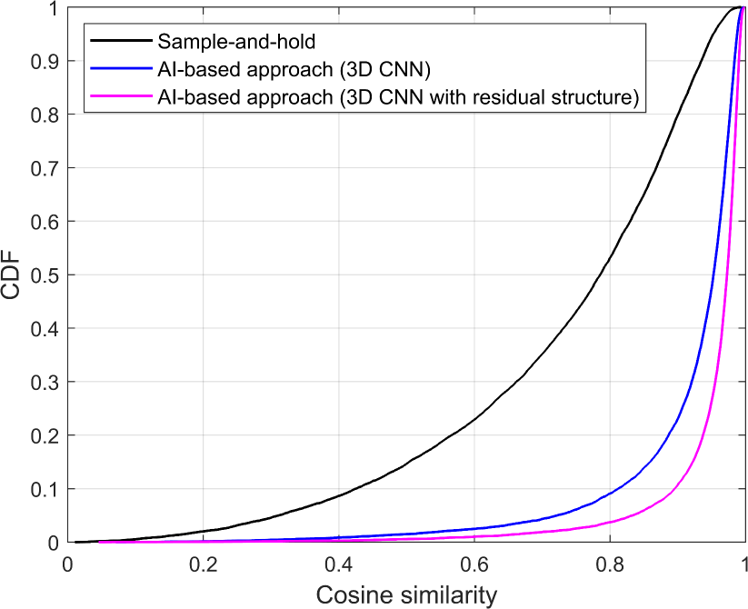}
	\caption{The CDF of the cosine-similarity performance illustrating the improvement brought by the residual structure.}
	\label{simu_fig_3}
\end{figure}

\begin{figure*}[t]
\centering
  \subfigure[30-50 km/h]{ \includegraphics[width=0.45\linewidth]{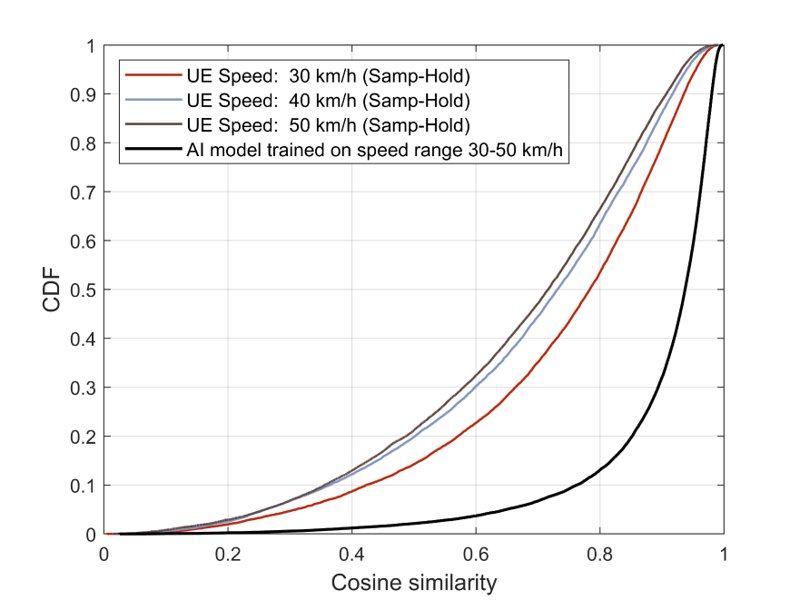} \label{simu_fig_4} }
  \subfigure[100-150 km/h]{ \includegraphics[width=0.45\linewidth]{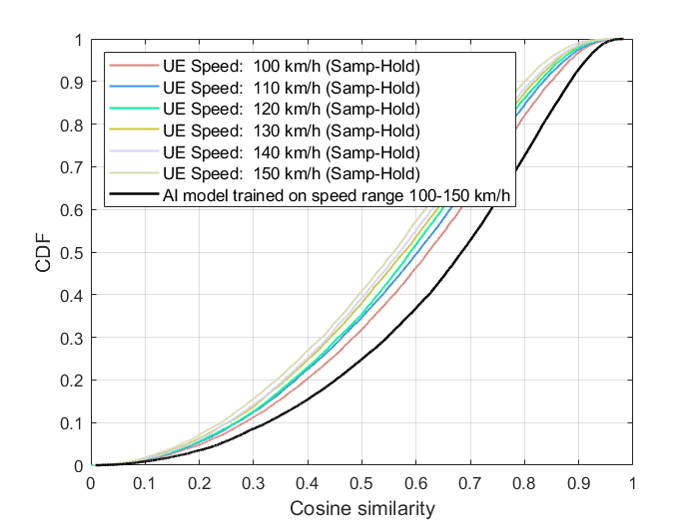} \label{simu_fig_5} }
  \caption{The CDF of the cosine-similarity performance of the proposed AI-based approach and the sample and hold approach under the speed range of (a) 30-50 km/h and (b) 100-150 km/h.}
  \label{simu_fig_4_5}
\end{figure*}

We first consider a UE that has a constant speed of $30$ km/h. \fref{simu_fig_1} illustrates the better performance of the deep learning model than sample and hold approach by plotting the heat map of the channel at each receive antenna, where each pixel represents the amplitude of the complex valued channel element from each transmit antenna and at each resource block. As can be seen, the neural network is able to capture most of the "hot spots" of the future channel. By contrast, due to the movement of the UE, the SH approach suffers from the channel aging and produces significant difference with respect to the ground true future channel.

In addition to the channel images, we further investigate the performance in terms of the cosine similarity with the ground true layer one beamforming vectors, which has direct influence on the achieved beamforming gain. As can be seen from \fref{simu_fig_2}, the AI-based approach achieves quite excellent performance, with most of the beamforming vectors calculated based on the predicted channels having over $0.9$ cosine similarity with the ground true layer one beamforming vectors. This forms direct contrast with the SH approach, where the performance of its predicted beamforming vectors is quite dispersive with non-negligible tails extending toward zero. And to be more specific, the AI-based prediction achieves $25.89\%$ improvement over sample and hold approach in terms of average cosine similarity with the true beamforming vectors.


Furthermore, we want to understand how much of the improvement does the residual structure bring to the system by learning the deeper patterns of the channel. We remove the two "Conv Res Blocks" shown in \fref{dl_model}, leaving only the two convolutional blocks, which is named as the 3D CNN model. As can be seen from \fref{simu_fig_3}, the performance of the 3D CNN model is also quite excellent compared to the sample and hold approach. However, by leveraging the powerful residual structure, it allows the model to learn finer details of the channel evolving pattern, which contributes another precious $3.6\%$ improvement.


\subsection{Results for the Mixed-Speed Dataset}

We further train and test the proposed deep learning model in two mixed-speed datasets, that is, a mid-speed dataset of 30-50 km/h and a high-speed dataset of 100-150 km/h. As can be seen from \fref{simu_fig_4_5}, the AI-based approach outperforms the SH approach in both cases, with a $29\%$ improvement at the mid-speed and a $14\%$ improvement at the high-speed. It is worth pointing out that the considered range at the high-speed is even larger than the mid-speed, which makes the channel prediction task even harder. Besides, we fix the number of past observed channels used for the deep learning model to predict the future channels in both cases. However, this parameter could be probably fine-tuned to further enhance the performance of the AI-based approach, which is left as future work.


%

\section{Conclusions and Discussions} \label{sec:Con}

In this paper, we develop a deep learning based channel prediction framework that proactively predicts the downlink channel state information based on the past observed channel sequence.
In its core, the model adopts a 3-D convolutional neural network (CNN) based architecture to efficiently learn the temporal, spatial and frequency correlations of downlink channel samples, based on which accurate channel prediction can be performed.
Simulation results highlight the potential of the developed learning model in extracting information and predicting future downlink channels directly from the observed past channel sequence, which significantly improves the performance compared to the sample-and-hold approach, and mitigates the impact of the dynamic communication environment.


\end{document}